\documentclass[preprint,12pt]{elsarticle}
\usepackage{amsmath}
\usepackage{amsfonts}
\usepackage{amssymb}
\usepackage{graphicx}
\usepackage{bm}
\numberwithin{equation}{section}
\begin{document}
\begin{frontmatter}
\title{The Wigner function in the relativistic quantum mechanics}
\author{K. Kowalski}
%\ead{kowalski@uni.lodz.pl}
%\cortext[cor1]{Corresponding author.}
\author{J. Rembieli\'nski}
\address{Department of Theoretical Physics, University
of \L\'od\'z, ul.\ Pomorska 149/153, 90-236 \L\'od\'z,
Poland}
\begin{abstract}
A detailed study is presented of the relativistic Wigner function for a 
quantum spinless particle evolving in time according to the Salpeter 
equation.
\end{abstract}
\begin{keyword}
Quantum mechanics, relativistic quantum mechanics; Wigner quasi-probability function; 
\end{keyword}
\end{frontmatter}
\section{Introduction}
The Wigner function also referred to as the Wigner quasi-probability 
function is one of the most important concepts of nonrelativistic quantum 
mechanics.  Its applications range from nonequilibrium quantum mechanics, 
quantum optics, quantum chaos and quantum computing to classical optics and 
signal processing.  As far as we are aware, in spite of the fact that the paper 
by Wigner was dated 1932 \cite{1}, the relativistic generalization of the 
Wigner function in the simplest case of the spinless particle was introduced by 
Zavialov and Malokostov only in 1999 \cite{2}.  The dynamics of that relativistic 
Wigner function was studied in recent papers \cite{3,4} by Larkin and Filonov.  
Clearly, the difficulties in extending the concept of the Wigner function 
to the relativistic domain sometimes considered as unattainable \cite{5} are 
closely related with problems in finding the relativistic counterpart of 
the Schr\"odinger equation, that is constructing the relativistic quantum 
mechanics.  In this work we discuss the advantages and limitations of the 
Wigner function introduced by Zavialov and Malokostov and analyze an 
alternative relativistic generalization of the Wigner function based on 
the standard nonrelativistic formula that was applied earlier in the case of the 
Dirac particle.  Both of these approaches utilize the relativistic quantum dynamics 
described by the spinless Salpeter equation.  The theory is illustrated by 
concrete examples of relativistic Wigner function for a spinless free 
particle.
\section{The Zavialov-Malokostov Wigner function}
\subsection{Definition of the Wigner function}
We now summarize the basic facts about the Wigner function introduced in 
ref.\ 2.  The point of departure in \cite{2} was the following form of the 
non-relativistic Wigner function
\begin{equation}
%<2.1>
W({\bm x},{\bm p},t)=\frac{1}{(2\pi)^3}\frac{1}{\hbar^6}\int d^3{\bm p}_1
d^3{\bm p}_2\tilde\phi^*({\bm p}_1,t)\tilde\phi({\bm p}_2,t)\delta({\bm p}
-{\textstyle{\frac{1}{2}}}({\bm p}_1+{\bm p}_2)) e^{\frac{{\rm i}({\bm p}_2-{\bm p}_1)
\mbox{\boldmath$\scriptstyle{\cdot}$}{\bm x}}{\hbar}}, 
\end{equation}
where $\tilde\phi({\bm p},t)$ is the Fourier transform of the wave function 
$\phi({\bm x},t)$, that is 
\begin{equation}
%<2.2>
\tilde\phi({\bm p},t)=\frac{1}{(2\pi)^\frac{3}{2}}\int d^3{\bm x}\,e^{-{\rm i}
\frac{{\bm p}\mbox{\boldmath$\scriptstyle{\cdot}$}{\bm x}}{\hbar}}\phi({\bm x},t).
\end{equation}
Furthermore, Zavialov and Malokostov restrict to the case of the free 
relativistic evolution described by the Salpeter equation (see \cite{6} and references
therein)
\begin{equation}
%<2.3>
{\rm i}\hbar\frac{\partial\tilde\phi({\bm p},t)}{\partial t}=\sqrt
{{\bm p}^2c^2+m^2c^4}\tilde\phi({\bm p},t),
\end{equation}
and demand that the relativistic Wigner function has the basic properties 
of the nonrelativistic one referring to integration over the spatial and 
momentum variables such that
\begin{align}
%<2.4>
\int d^3{\bm p}W({\bm x},{\bm p},t) &= |\phi({\bm x},t)|^2=\rho({\bm 
x},t),\\
\int d^3{\bm x}W({\bm x},{\bm p},t) &= \frac{1}{\hbar^3}|\tilde\phi({\bm p},t)|^2
=\rho_p({\bm p},t),
\end{align}
and satisfy the evolution law
\begin{equation}
%<2.6>
W({\bm x},{\bm p},t+\tau)=W({\bm x}-\frac{c{\bm p}}{p_0}\tau,{\bm p},t),
\end{equation}
where $p_0=E/c=\sqrt{{\bm p}^2+m^2c^2}$.  It is easy to verify that the 
evolution law (2.6) is the global form of the local relation
\begin{equation}
%<2.7>
\frac{\partial W({\bm x},{\bm p},t)}{\partial t}+\frac{c{\bm p}}{p_0}{\bm\cdot}{\bm\nabla}
W({\bm x},{\bm p},t)=0
\end{equation}
generalizing the nonrelativistic equation that is valid in the case of the free
evolution 
\begin{equation}
%<2.8>
\frac{\partial W({\bm x},{\bm p},t)}{\partial t}+\frac{{\bm p}}{m}{\bm\cdot}{\bm\nabla}
W({\bm x},{\bm p},t)=0.
\end{equation}
With these assumptions the following relativistic generalization of (2.1) 
was obtained in \cite{2}:
\begin{equation}
%<2.9>
W({\bm x},{\bm p},t)=\frac{1}{(2\pi)^3}\frac{1}{\hbar^6}\int d^3{\bm p}_1
d^3{\bm p}_2\tilde\phi^*({\bm p}_1,t)\tilde\phi({\bm p}_2,t)\delta({\bm p}
-({\bm p}_1\oplus{\bm p}_2)) e^{\frac{{\rm i}({\bm p}_2-{\bm p}_1)
\mbox{\boldmath$\scriptstyle{\cdot}$}{\bm x}}{\hbar}}, 
\end{equation}
where ${\bm p}_1\oplus{\bm p}_2$ is a counterpart of the sum on the mass 
hyperboloid.  More precisely, ${\bm p}_1\oplus{\bm p}_2$ is the spacial 
part of the fourvector $p_1\oplus p_2$ on the mass hyperboloid $p^2=m^2c^2$
of the form
\begin{equation}
%<2.10>
p_1\oplus p_2=mc\frac{p_1+p_2}{\sqrt{(p_1+p_2)^2}},
\end{equation}
so $(p_1\oplus p_2)^2=m^2c^2$, and is given by
\begin{equation}
%<2.11>
{\bm p}_1\oplus{\bm p}_2=mc\frac{{\bm p}_1+{\bm p}_2}{\sqrt{2(m^2c^2+
p_0({\bm p}_1)p_0({\bm p}_2)-{\bm 
p}_1\mbox{\boldmath$\scriptstyle{\cdot}$}{\bm p}_2})},
\end{equation}
where $p_0({\bm p}_i)=\sqrt{{\bm p}_i^2+m^2c^2}$, $i=1,\,2$.
The formula (2.9) can be immediately generalized to involve the particle 
in the external potential $V({\bm x})$ \cite{4} by demanding that $\tilde\phi({\bm x},t)$
in (2.9) fulfils the Salpeter equation in the momentum representation 
\begin{equation}
%<2.12>
{\rm i}\hbar\frac{\partial\tilde\phi({\bm p},t)}{\partial t}=[\sqrt
{m^2c^4+{\bm p}^2c^2} + V({\rm i}\hbar{\bm\nabla}_{\bm p})]\tilde\phi({\bm p},t).
\end{equation}
\subsection{The Wigner function and probability current}
An interesting property of the relativistic Wigner function (2.9) that was 
not recognized neither in \cite{2} nor \cite{3,4} is the following easily proven 
relation
\begin{equation}
%<2.13>
{\bm j}({\bm x},t)=\int d^3{\bm p}\frac{c{\bm p}}{p_0}W({\bm x},{\bm p},t),
\end{equation}
where ${\bm j}({\bm x},t)$ is the relativistic probability current 
introduced by us in ref. 6, describing the conservation of the probability 
in the Salpeter equation in the coordinate representation
\begin{equation}
%<2.14>
{\rm i}\hbar\frac{\partial\phi({\bm x},t)}{\partial
t}=[\sqrt{m^2c^4-\hbar^2c^2\Delta}+V({\bm x})]\phi({\bm x},t),
\end{equation}
where $\Delta\equiv{\bm\nabla^2}$, via the continuity equation
\begin{equation}
%<2.15>
\frac{\partial\rho}{\partial t} + {\bm\nabla}{\bm\cdot}{\bm j} = 0,
\end{equation}
where $\rho({\bm x},t)=|\phi({\bm x},t)|^2$ is the probability density,
such that
\begin{equation}
%<2.16>
{\bm j}({\bm x},t) = \frac{c}{(2\pi)^3\hbar^6}\int d^3{\bm p}d^3
{\bm k}\,\frac{{\bm p}+{\bm k}}{p_0({\bm p})+p_0({\bm k})}\,e^{{\rm i}\frac{({\bm k}-{\bm p})
\mbox{\boldmath$\scriptstyle{\cdot}$}{\bm x}}{\hbar}}
\tilde\phi^*({\bm p},t)\tilde\phi({\bm k},t).
\end{equation}
Of course, (2.13) is the relativistic generalization of the relation
\begin{equation}
%<2.17>
{\bm j}({\bm x},t)=\int d^3{\bm p}\frac{{\bm p}}{m}W({\bm x},{\bm p},t),
\end{equation}
leading via the formula for quantum expectation values of an observable 
$\hat A$
\begin{equation}
%<2.18>
\langle\phi|\hat A\phi\rangle=\int d^3{\bm x}d^3{\bm p}A({\bm x},{\bm 
p})W({\bm x},{\bm p}),
\end{equation}
where $A({\bm x},{\bm p})$ is the Weyl transform of the operator $\hat A$,
to the well-known expression describing the connection of the integral of the 
probability current and average velocity in the given state 
\begin{equation}
%<2.19>
\int{\bm j}({\bm x},t)\,d^3{\bm x} = \langle\phi|\hat{\bm v}\phi\rangle,
\end{equation}
where $\hat{\bm v}={\bm p}/m$ is the velocity operator.  It must be borne 
in mind that (2.13) cannot be regarded as the definition of the relativistic 
probability current.  The proper definition of the probability current that holds
regardless of the accepted definition of the Wigner function and ensures 
the validity of the continuity equation (2.15) is (2.16).  Nonetheless, the
formula (2.13) is really remarkable.
\subsection{Massless limit of the Wigner function}
As we have seen the relativistic Wigner function (2.9) has some nice 
properties as (2.4), (2.5), (2.7) and (2.13).  Nevertheless, the authors 
of \cite{2} and \cite{3,4} seem to be unaware of its problematic behavior in the 
limit $m=0$.  Indeed consider for simplicity the case of a free relativistic particle 
on a line.  The relativistic Wigner function takes then the form
\begin{equation}
%<2.20>
W(x,p,t)=\frac{1}{2\pi\hbar^2}\int dp_1dp_2\tilde\phi^*(p_1,t)
\tilde\phi(p_2,t)\delta(p-(p_1\oplus p_2))e^{\frac{{\rm i}(p_2-p_1)x}{\hbar}},
\end{equation}
where $\tilde\phi(p,t)$ satisfies the Salpeter equation
\begin{equation}
%<2.21>
{\rm i}\hbar\frac{\partial\tilde\phi(p,t)}{\partial t}=\sqrt
{p^2c^2+m^2c^4}\,\tilde\phi(p,t), 
\end{equation}
and
\begin{equation}
%<2.22>
p_1\oplus p_2=mc\frac{p_1+p_2}{\sqrt{2(m^2c^2+p_0(p_1)p_0(p_2)-p_1p_2)}},
\end{equation}
where $p_0(p_i)=\sqrt{p_i^2+m^2c^2}$, $i=1,\,2$.  Now we have the parametric form of the 
Wigner function introduced in ref.\cite{2} that can be easily obtained from (2.20) by switching 
to coordinates $p_{1,2}=mc\sinh\gamma_{1,2}$ on the mass-shell hyperboloid:
\begin{align}
%<2.23>
&W(x,p,t)\nonumber\\
&=\frac{mc}{\pi\hbar^2}\frac{1}{\cosh\kappa}\int_{-\infty}^\infty
d\beta\cosh(\kappa+\beta)\cosh(\kappa-\beta)\tilde\phi^*(mc\sinh(\kappa+\beta),t)\nonumber\\
&\hspace*{1em}\times\tilde\phi(mc\sinh(\kappa-\beta),t)\exp\left\{\frac{{\rm i}mc}{\hbar}
[\sinh(\kappa-\beta)-\sinh(\kappa+\beta)]x\right\}, 
\end{align}
where $p=mc\sinh\kappa$. From (2.23) we can derive the following formula for the Wigner function
\begin{align}
%<2.24>
&W(x,p,t)=\frac{1}{\pi\hbar^2\sqrt{p^2+m^2c^2}}\nonumber\\
&\hspace*{1em}\times\int_{-\infty}^\infty 
d\beta(p^2+m^2c^2\cosh^2\beta)\tilde\phi^*(p\cosh\beta+\sqrt{p^2+m^2c^2}\sinh\beta,t)\nonumber\\
&\hspace*{1em}\times\tilde\phi(p\cosh\beta-\sqrt{p^2+m^2c^2}\sinh\beta,t)
\exp\left(-\frac{2{\rm i}x}{\hbar}\sqrt{p^2+m^2c^2}\sinh\beta\right).
\end{align}
An immediate consequence of (2.24) is the massless limit of the Wigner 
function such that
\begin{align}
%<2.25>
&W_0(x,p,t)=\lim_{m\to0}W(x,p,t)\nonumber\\
&=\frac{|p|}{\pi\hbar^2}\int_{-\infty}^\infty 
d\beta\tilde\phi^*(p\cosh\beta+|p|\sinh\beta,t)\tilde\phi(p\cosh\beta-|p|\sinh\beta,t)
e^{-\frac{2{\rm i}x}{\hbar}|p|\sinh\beta},
\end{align}
where $\tilde\phi(p,t)$ fulfills the Salpeter equation
\begin{equation}
%<2.26>
{\rm i}\hbar\frac{\partial\tilde\phi(p,t)}{\partial t}=c|p|\tilde\phi(p,t). 
\end{equation}
We point out that for a spinless particle we have no problems connected with
procedures of contractions of representations of little groups corresponding
to massive and massless particles.  Indeed, we then deal in both cases with trivial
representations. The limit (2.25) gives the correct density in the momentum 
representation i.e.
\begin{equation}
%<2.27>
\int_{-\infty}^\infty W_0(x,p,t)dx=\frac{1}{\hbar}|\tilde\phi(p,t)|^2.
\end{equation}
Nevertheless, it leads to erroneous formula for the density in the 
coordinate representation and the probability current.  We now illustrate 
this observation by the example of the ``Lorentzian" wave packet \cite{6}.

Consider the following normalized solution \cite{6}
\begin{equation}
%<2.28>
\tilde\phi(p,t) = \sqrt{a}e^{-(a+{\rm i}t)|p|},
\end{equation}
where $a>0$, to the Salpeter equation in the momentum representation for a 
massless particle moving in a line (2.26), where we set $\hbar=1$ and 
$c=1$.  The normalized wave function corresponding to (2.28) satisfying the 
Salpeter equation
\begin{equation}
%<2.29>
{\rm i}\frac{\partial\phi(x,t)}{\partial t}=\sqrt{-\frac{\partial^2}{\partial x^2}}\,\phi(x,t),
\end{equation}
is given by
\begin{equation}
%<2.30>
\phi(x,t)=\sqrt{\frac{2a}{\pi}}\frac{a+{\rm i}t}{x^2+(a+{\rm i}t)^2}.
\end{equation}
From (2.30) and the one-dimensional counterpart of (2.16) for $m=0$ such 
that
\begin{equation}
%<2.31>
j(x,t) = \frac{1}{2\pi}\int dpdk\,\frac{p+k}{|p|+|k|}e^{{\rm i}(k-p)x}\tilde\phi^*(p,t)
\tilde\phi(k,t),
\end{equation}
we immediately get the following formulas for the probability density and 
current, respectively \cite{6}
\begin{align}
%<2.32>
\rho(x,t)&=|\phi(x,t)|^2=\frac{2a}{\pi}\frac{a^2+t^2}{(x^2-t^2+a^2)^2+4a^2t^2},\\
j(x,t)&= \frac{a}{4\pi t^2}\ln\frac{(x+t)^2+a^2}{(x-t)^2+a^2}-\frac{ax}{\pi t}
\frac{x^2-3t^2+a^2}{(x^2-t^2+a^2)^2+4a^2t^2}.
\end{align}
We now return to (2.25).  Inserting (2.28) into (2.25) where we set 
$\hbar=1$, and $c=1$, we find that the probability density and probability 
current corresponding to the massless limit are expressed by
\begin{align}
%<2.34>
\rho_0(x,t)&=\int_{-\infty}^\infty W_0(x,p,t)dp=\frac{a}{2\pi}\left[\frac{1}{(x-t)^2+a^2}
+\frac{1}{(x+t)^2+a^2}\right],\\
j_0(x,t)&=\int_{-\infty}^\infty\frac{p}{|p|} W_0(x,p,t)dp=\frac{a}{2\pi}\left[\frac{1}{(x-t)^2+a^2}
-\frac{1}{(x+t)^2+a^2}\right].
\end{align}
Thus it turns out that $\rho_0$ and $j_0$ obtained from the massless limit 
of the Wigner function are different from the correct probability density 
$\rho$ and probability current $j$ given by (2.32) and (2.33), 
respectively.  We point out that $\rho_0$ and $j_0$ can be written as
\begin{align}
%<2.36>
\rho_0(x,t)&=\frac{1}{2}(|\phi_+(x,t)|^2+|\phi_-(x,t)|^2),\\
j_0(x,t)&=\frac{1}{2}[j_+(x,t)+j_-(x,t)],
\end{align}
where
\begin{equation}
%<2.38>
\rho_\pm(x,t) = |\phi_\pm(x,t)|^2 = \pm j_\pm(x,t)=
\frac{a}{\pi}\frac{1}{(x\mp t)^2+a^2},
\end{equation}
and $\rho_\pm$ is the probability density and $j_\pm$ is the probability 
current related to the wave packet $\phi_\pm$ referring to the particle 
moving to the right and left, respectively such that \cite{6}
\begin{align}
%<2.39>
\phi_\pm(x,t)&= \sqrt{\frac{a}{\pi}}\frac{\pm{\rm i}}{x\mp t\pm {\rm 
i}a},\\
\rho_\pm(x,t)&= |\phi_\pm(x,t)|^2 = \pm j_\pm(x,t)=
\frac{a}{\pi}\frac{1}{(x\mp t)^2+a^2},
\end{align}
where 
\begin{equation}
%<2.41>
\tilde\phi_\pm(p,t)=\sqrt{2a}\theta(\pm p)e^{-(a+{\rm i}t)|p|}
\end{equation}
and $\theta(p)$ is the Heaviside step function.  Since 
\begin{equation}
%<2.42>
\phi(x,t)=\frac{1}{\sqrt{2}}[\phi_+(x,t)+\phi_-(x,t)],
\end{equation}
where $\phi(x,t)$ is given by (2.30), therefore $\rho_0(x,t)$ differs from 
the correct probability density in lack of the interference terms.  We 
remark that $\rho_0(x,t)$ and $j_0(x,t)$ satisfy the continuity 
equation
\begin{equation}
%<2.43>
\frac{\partial\rho_0}{\partial t}+\frac{\partial j_0}{\partial x}=0.
\end{equation}

In spite of their suggestive form, the problems with the massless limit of 
the relativistic Wigner function (2.9) are related to the conditions (2.7) and (2.13).
To see this let us assume that (2.7) holds in the ultra-relativistic limit
$m\to0$.  Furthermore, we confine to the particle on a line, so we have
\begin{equation}
%<2.44>
\frac{\partial W(x,p,t)}{\partial t}+\frac{cp}{|p|}\frac{\partial W(x,p,t)}
{\partial x}=0.
\end{equation}
Consider now the relation
\begin{equation}
%<2.45>
j(x,t)=\int_{-\infty}^\infty dp\frac{cp}{|p|}W(x,p,t),
\end{equation}
that is the massless limit of (2.13).  Using (2.44) we get
\begin{equation}
%<2.46>
\frac{\partial j(x,t)}{\partial t}=-c^2\frac{\partial}{\partial x}
\int_{-\infty}^\infty dpW(x,p,t)=-c^2\frac{\partial\rho(x,t)}{\partial x},
\end{equation}
where $\rho(x,t)=|\phi(x,t)|^2$ is the probability density.  Combining 
this with the continuity equation
\begin{equation}
%<2.47>
\frac{\partial\rho(x,t)}{\partial t}+\frac{\partial j(x,t)}{\partial x}=0,
\end{equation}
we arrive at the wave equation
\begin{equation}
%<2.48>
\frac{1}{c^2}\frac{\partial^2\rho(x,t)}{\partial t^2}-\frac{\partial^2\rho(x,t)}
{\partial x^2}=0.
\end{equation}
On the contrary, in view of (2.32) and (2.33) both relations (2.46) and so 
(2.48) are easily shown to be erroneous in the case with the free 
evolution of the massless particle described by the wavefunction (2.30).

The formula for the Wigner function referring to the solution (2.28) can 
be derived with the help of (2.25) where we set $\hbar=1$, and the 
identity (see \cite{7} and \cite{8})
\begin{equation}
%<2.49>
\int_0^\infty dx\,\frac{\exp(-\alpha\sqrt{x^2+\beta^2})}{\sqrt{x^2+\beta^2}}
\cos\gamma x =K_0(\beta\sqrt{\alpha^2+\gamma^2}),\qquad{\rm Re}\alpha >0,\, 
{\rm Re}\beta >0.
\end{equation}
Namely, we have
\begin{equation}
%<2.50>
W_0(x,p,t)=\frac{2a|p|}{\pi}K_0[2\sqrt{a^2p^2+(pt-x|p|)^2}],
\end{equation}
where $K_\nu(z)$ is the modified Bessel function (Macdonald function).  
We remark that the Wigner function (2.50) is nonnegative.  The time 
evolution of the Wigner function is shown in Fig.\ 1.  As one might expect 
in view of the form of the wave packet (2.42) we have two stable maxima of 
the quasiprobability function (2.50) --- one moving to the left and one 
moving to the right with the same constant absolute value of the momenta.
\begin{figure*}
%<figure 1>
\centering
\begin{tabular}{c@{}c}
\includegraphics[scale=.7]{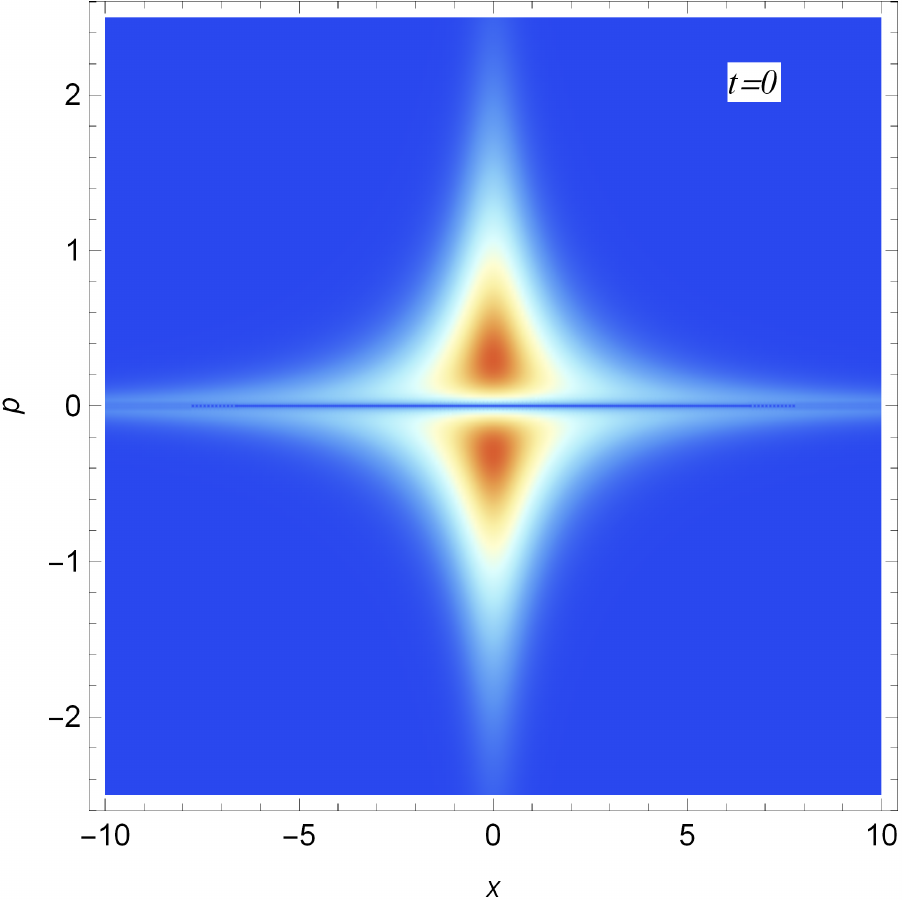}&
\includegraphics[scale=.7]{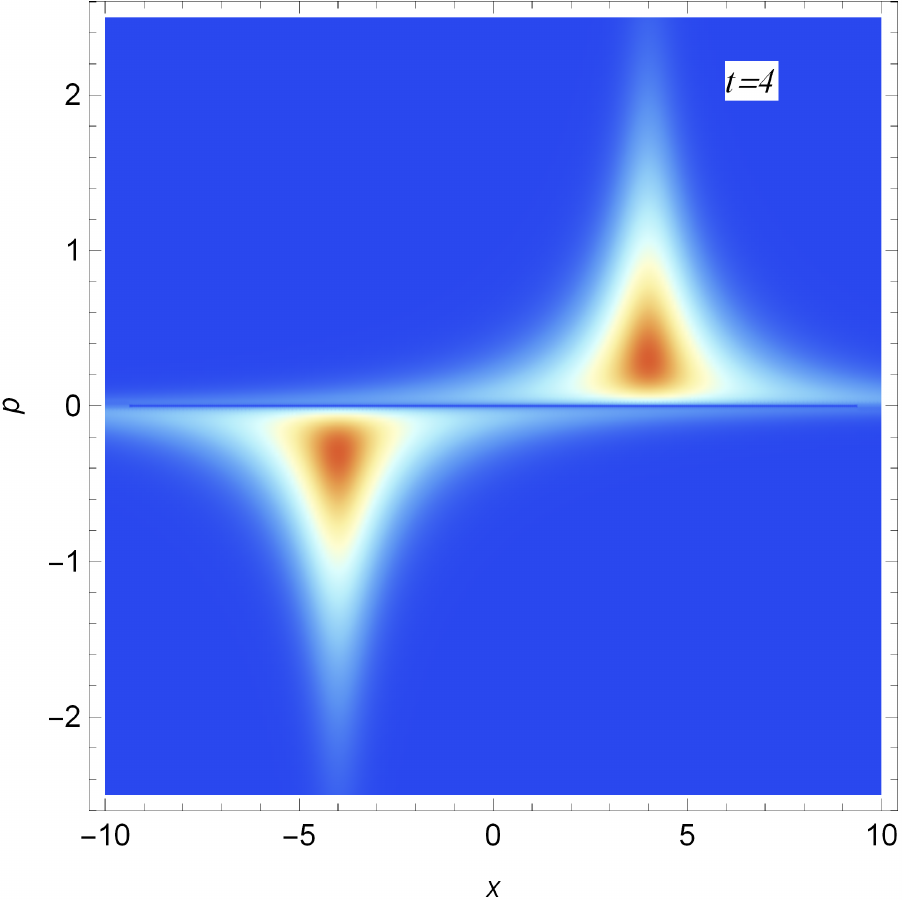}\\
\end{tabular}
\caption{The density plot illustrating the time evolution of the Wigner 
function (2.50) corresponding to the solution of the Salpeter equation given 
by the wave packet for a free massless particle on a line (2.31). The 
parameter $a=1$.  The stable maxima refer to the particle moving to the left
and to the right described by (2.39).}
\end{figure*}

We now discuss the solutions $\phi_\pm(x,t)$ to the Salpeter equation 
(2.29) given by (2.39) referring to the particle moving to the right and 
left.  Using (2.41) and proceeding as with (2.28) we get the following 
Wigner function (2.25) with $\hbar=1$, corresponding to $\phi_\pm(x,t)$, 
respectively
\begin{equation}
%<2.51>
W_{0\pm}(x,p,t)=\pm\theta(\pm p)\frac{4ap}{\pi}K_0[\pm2p\sqrt{(x\mp t)^2+a^2}].
\end{equation}
As with (2.50) both Wigner functions (2.51) are nonnegative.
Of course, the plot of the function $W_{0\pm}(x,p,t)$ refers to the upper 
(lower) part of Fig.\ 1.  As mentioned earlier the functions $\phi_\pm(x,t)$
satisfy the condition (2.27), where we set $\hbar=1$.  Moreover, using the identity
\cite{8}
\begin{align}
%<2.52>
\int_0^\infty x^\mu 
K_\nu(ax)dx&=2^{\mu-1}a^{-\mu-1}\Gamma\left(\frac{1+\mu+\nu}{2}\right)
\Gamma\left(\frac{1+\mu-\nu}{2}\right),\nonumber\\
&\qquad{\rm Re}(\mu+1\pm\nu)>0,\quad {\rm Re}\,a>0,
\end{align} 
we find that, in opposition to (2.50), the Wigner functions (2.51) give 
the correct formulas for the density in the coordinate representation, that 
is we have
\begin{equation}
%<2.53>
\int_{-\infty}^\infty dp\,W_{0\pm}(x,p,t)=|\phi_\pm(x,t)|^2.
\end{equation}
We conclude that the Wigner functions corresponding to the wave packets 
$\phi_\pm(x,t)$ referring to the massless particle moving to the right and 
left respectively, satisfy all requirements imposed on the Wigner function 
valid in the massive case.  It seems that such good behavior of the Wigner 
functions (2.51) and bad one of the Wigner function (2.50) are 
related to the fact that in the massless case one can define the sum of 
vectors on a cone $p\to|p|$ that remains on a cone, analogous to (2.10), 
only for vectors with the same direction (proportional ones).  We finally 
remark that in opposition to the nonrelativistic case when Hudson theorem 
\cite{9} holds, which states that the only wave packet with non-negative 
Wigner function is the exponential of a quadratic polynomial, the wave
functions (2.39) corresponding to the non-negative Wigner functions (2.51)
are rational.  This is to the best of our knowledge, the first example in 
the literature of such wave packets for a relativistic spinless particle. 
For the relativistic spin one-half particle the wave functions violating the 
Hudson theorem were introduced recently in ref.\cite{10}.
\subsection{The Wigner function for the massive case}
We finally discuss the normalized solution \cite{6}
\begin{equation}
%<2.54>
\tilde\phi(p,t) = \frac{1}{\sqrt{2mK_1(2ma)}}e^{-(a+{\rm 
i}t)\sqrt{p^2+m^2}},
\end{equation}
where $a>0$, to the Salpeter equation in the momentum representation for a 
free massive particle moving in a line
\begin{equation}
%<2.55>
{\rm i}\frac{\partial\tilde\phi(p,t)}{\partial 
t}=\sqrt{m^2+p^2}\,\tilde\phi(p,t),
\end{equation} 
where we set $\hbar=1$ and $c=1$.  The solution (2.54) is the Fourier 
transform of the following normalized wave function \cite{6}
\begin{equation}
%<2.56>
\phi(x,t) = \sqrt{\frac{m}{\pi K_1(2ma)}}\frac{a+{\rm
i}t}{\sqrt{x^2+(a+{\rm i}t)^2}}K_1[m\sqrt{x^2+(a+{\rm i}t)^2}]
\end{equation}
satisfying the Salpeter equation of the form
\begin{equation}
%<2.57>
{\rm i}\frac{\partial\phi(x,t)}{\partial t} =
\sqrt{m^2-\frac{\partial^2}{\partial x^2}}\,\phi(x,t).
\end{equation}
It can be easily demonstrated with the help of the asymptotic formula
\begin{equation}
%<2.58>
K_1(z) = \frac{1}{z},\qquad z\to 0,
\end{equation}
that (2.28) and (2.30) are the massless limits of (2.54) and (2.56), 
respectively.  In this sense (2.54) and (2.56) are the massive 
generalizations of (2.28) and (2.30).  Now, taking into account the 
definition in the parametric form (2.23), the identity (2.49) and 
elementary properties of the modified Bessel function we get after some 
calculation the following formula for the Wigner function corresponding to 
(2.54)
\begin{align}
%<2.59>
W(x,p,t)=&\frac{m}{2\pi 
K_1(2ma)}\frac{1}{\sqrt{p^2+m^2}}\nonumber\\
&\times\left\{
\left(1+\frac{2p^2}{m^2}\right)K_0\left[2\sqrt{a^2(p^2+m^2)+(tp-x\sqrt{p^2+m^2})^2}\right]
\right.\nonumber\\
&{}+\frac{a^2(p^2+m^2)-(tp-x\sqrt{p^2+m^2})^2}{a^2(p^2+m^2)+(tp-x\sqrt{p^2+m^2})^2}\nonumber\\
&\times\left. K_2\left[2\sqrt{a^2(p^2+m^2)+(tp-x\sqrt{p^2+m^2})^2}\right]\right\}.
\end{align}
Using (2.58) we can verify that the Wigner function obtained for the 
massless particle (2.50) is indeed the limit $m\to0$ of (2.59), i.e.\ we 
have
\begin{equation}
%<2.60>
\lim_{m\to0}W(x,p,t)=W_0(x,p,t)=\frac{2a|p|}{\pi}K_0[2\sqrt{a^2p^2+(pt-x|p|)^2}].
\end{equation}
The time development of the Wigner function (2.59) is depicted in Fig.\ 2. 
\begin{figure*}
%<figure 2>
\centering
\begin{tabular}{c@{}c}
\includegraphics[width =.5\textwidth]{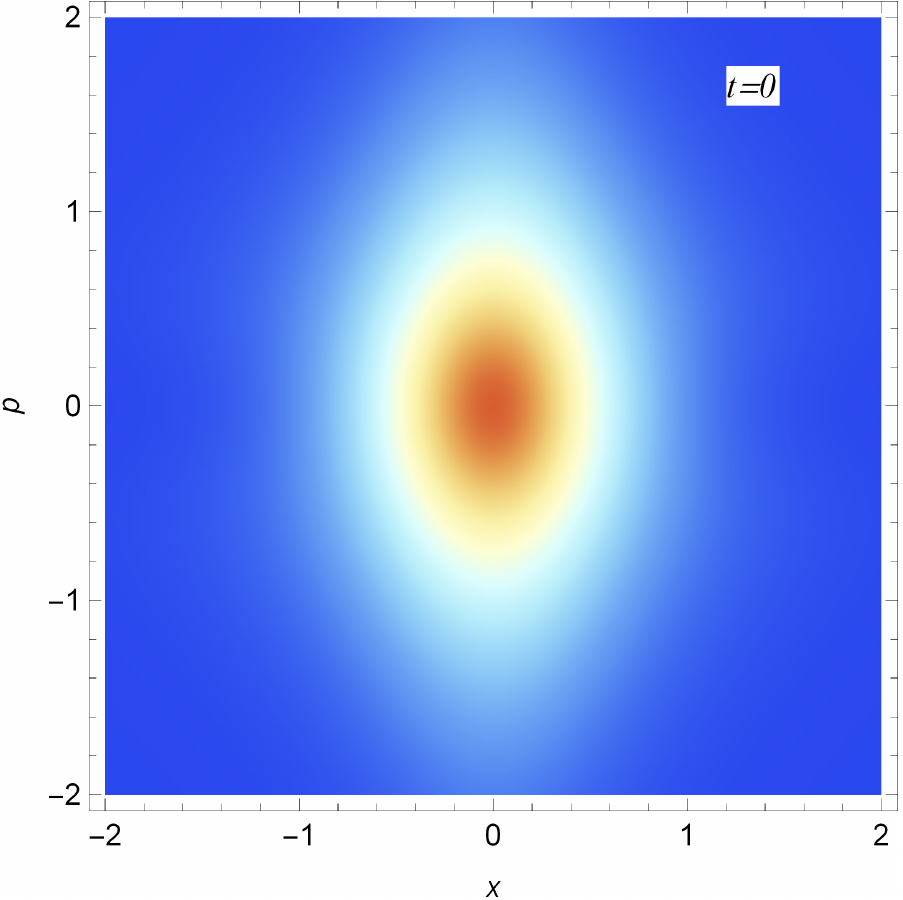}&
\includegraphics[width =.5\textwidth]{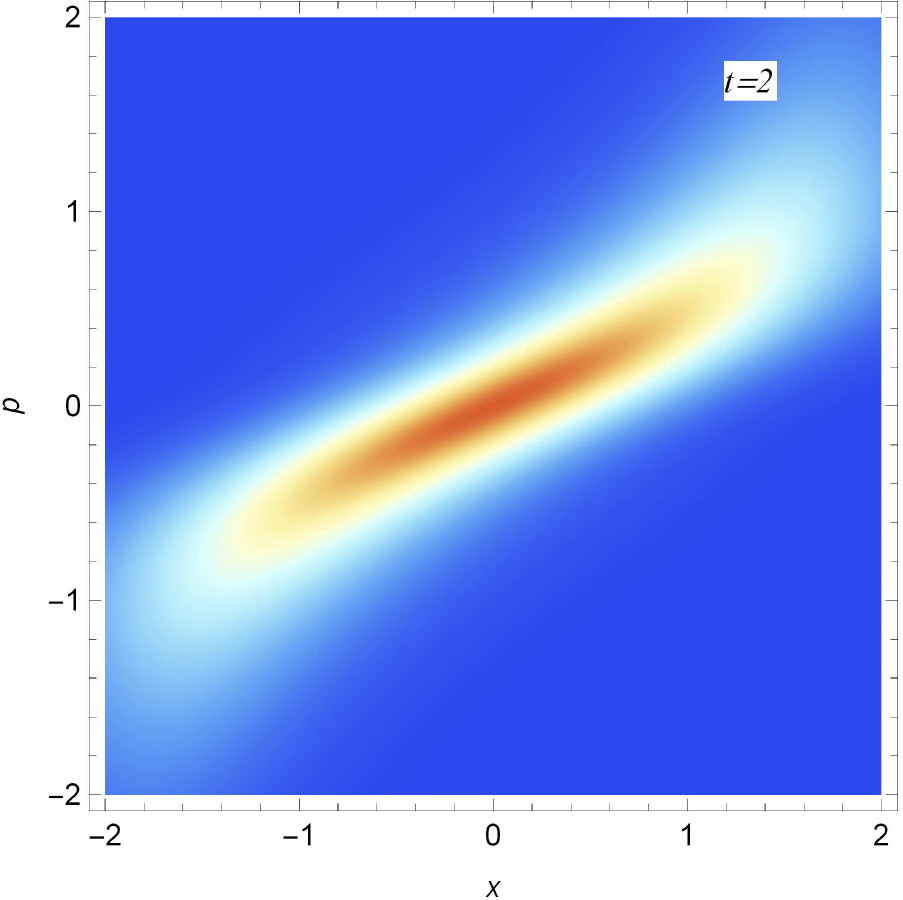}\\
\includegraphics[width =.5\textwidth]{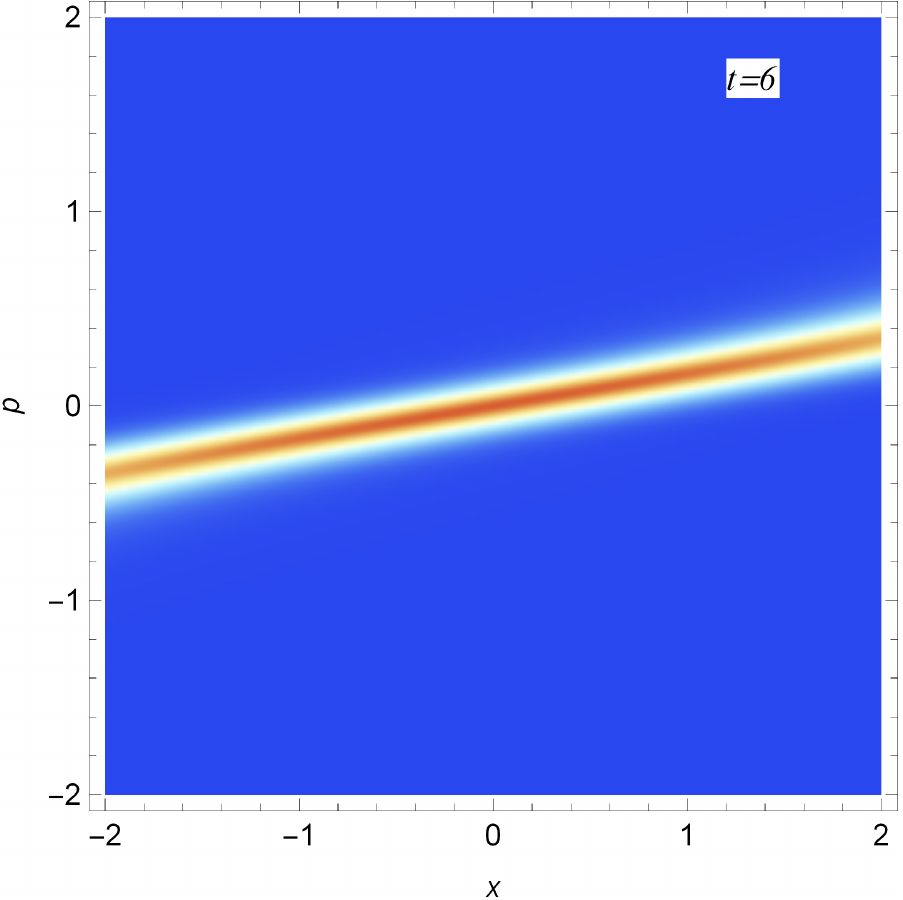}&
\includegraphics[width =.5\textwidth]{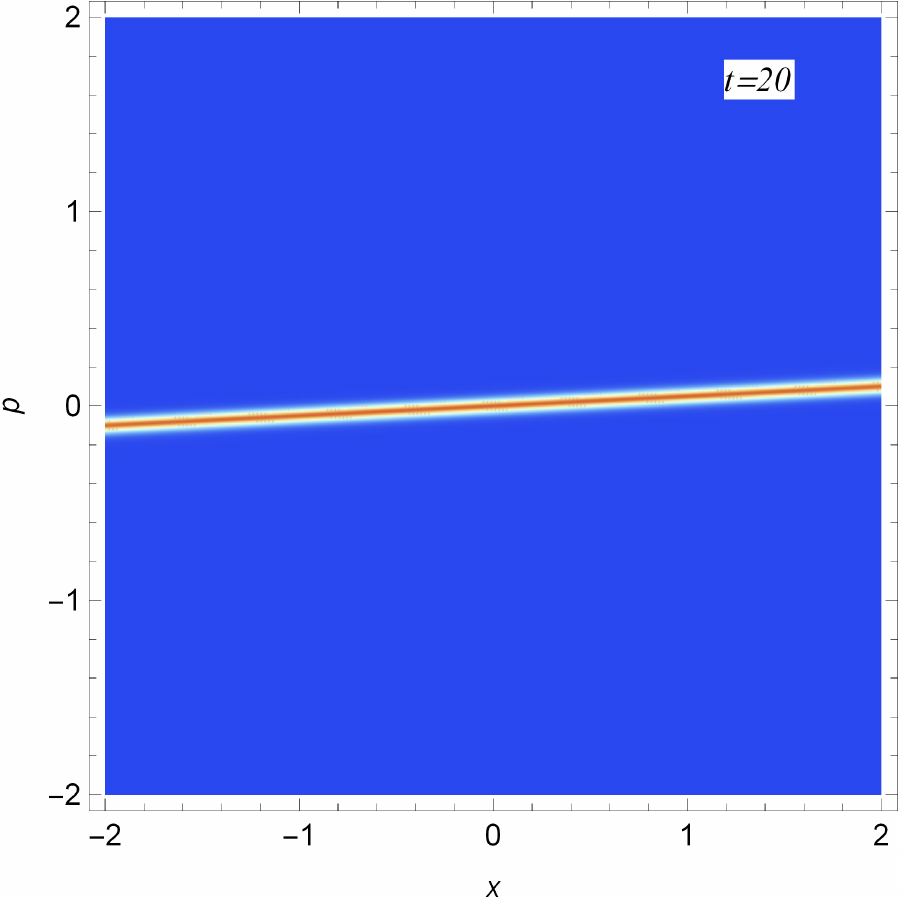}
\end{tabular}
\caption{The time development of the Wigner function (2.59) referring to the case
of a free massive particle on a line described by the solution to the Salpeter equation
in the momentum representation (2.54). The mass $m=1$ and $a=1$.}
\end{figure*}
We remark that for a wide range of parameters it is qualitatively similar to the time evolution 
of the nonrelativistic Wigner function for the (normalized) state that can be regarded as a 
counterpart of (2.54)
\begin{equation}
%<2.61>
\tilde\phi_\text{nrel}(p,t)=\left(\frac{a}{\pi m}\right)^{1/4}e^{-(a+{\rm i}t)\frac{p^2}{2m}},
\end{equation}
where $a>0$, such that
\begin{equation}
%<2.62>
W_\text{nrel}(x,p,t)=\frac{1}{\pi}\exp\left[-a\frac{p^2}{m}-\frac{m}{a}
\left(x-\frac{p}{m}t\right)^2\right].
\end{equation}
In particular, for large $t$ the Wigner function is concentrated around 
$p=0$ and goes to the uniform distribution along $x$-axis corresponding to 
its maximum.  Nevertheless, as easily seen from (2.59) by considering the 
case of small $a$ and $p$, in opposition to the nonrelativistic case 
(2.62), the Wigner function (2.59) can take negative values.  Such 
behavior is depicted in Fig.\ 3.  We also point out that, in contrast to  
the nonrelativistic Wigner function, the function (2.59) is not constrained
to be bounded via the inequality 
\begin{equation}
%<2.63>
|W(x,p,t)|\le\frac{2}{h},
\end{equation}
where $h$ is the Planck constant, that is a reflection of the uncertainty principle.
On the other hand, it is unclear what is the form of the upper bound in the general
case of the Wigner function (2.24).
\begin{figure*}
%<figure 3>
\centering
\includegraphics[scale=1]{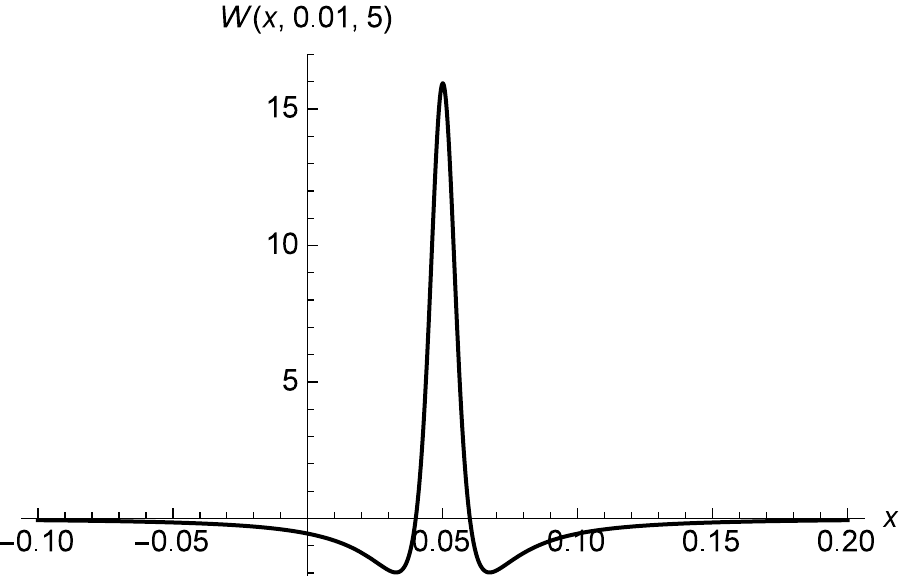}
\caption{The plot of the Wigner function $W(x,p,t)$ given by (2.59) with fixed 
$p=0.01$ and $t=5$ illustrating its nonpositivity.  The parameter $a=0.01$ and $m=1$.}
\end{figure*}

We finally write down the following equivalent form of the Wigner function 
(2.24) that can be regarded as a relativistic generalization of the 
standard formula for the Wigner function in the momentum space (see (3.14) 
in the next section)
\begin{align}
%<2.64>
W(x,p,t)=\frac{1}{4\pi\hbar^2}\int_{-\infty}^\infty dk&\frac{m^2c^2k^2+4(p^2+m^2c^2)^2}
{(p^2+m^2c^2)^{\frac{3}{2}}\sqrt{k^2+4(p^2+m^2c^2)}}\tilde\phi^*\left(\frac{f(p,k)-k}{2},t
\right)\nonumber\\
&\times\tilde\phi\left(\frac{f(p,k)+k}{2},t\right)e^{\frac{{\rm i}kx}{\hbar}},
\end{align}
where
\begin{equation}
%<2.65>
f(p,k)=\frac{p\sqrt{k^2+4(p^2+m^2c^2)}}{\sqrt{p^2+m^2c^2}}.
\end{equation}
Hence, taking the limit $m\to0$, we get the counterpart of (2.25) 
\begin{align}
%<2.66>
W_0(x,p,t)=\frac{|p|}{\pi\hbar^2}\int_{-\infty}^\infty&\frac{dk}{\sqrt{k^2+4p^2}}
\tilde\phi^*\left(\frac{\frac{p}{|p|}\sqrt{k^2+4p^2}-k}{2},t\right)\nonumber\\
&\times\tilde\phi\left(\frac{\frac{p}{|p|}\sqrt{k^2+4p^2}+k}{2},t\right)e^{\frac{{\rm i}kx}{\hbar}}.
\end{align}
the relation (2.64) can be obtained from (2.24) by formal substitution
\begin{equation}
%<2.67>
\sqrt{p^2+m^2c^2}\sinh\beta=-k/2.
\end{equation}
Nevertheless, it can be also obtained from (2.20) by effective integration 
of the delta function, without usage of coordinates on the mass-shell 
hyperboloid and the formula (2.24).

\section{The relativistic Wigner function based on the standard definition}
In our opinion the plausible relativistic generalization of the Wigner 
function for a free spinless particle is given by
\begin{equation}
%<3.1>
W({\bm x},{\bm p},t)=\frac{1}{(2\pi\hbar)^3}\int d^3{\bm q}\phi^*({\bm x}-{\bm q}/2,t)
\phi({\bm x}+{\bm q}/2,t)e^{-{\rm i}\frac{{\bm 
p}\mbox{\boldmath$\scriptstyle{\cdot}$}{\bm q}}{\hbar}},
\end{equation}
where $\phi({\bm x},t)$ satisfies the Salpeter equation
\begin{equation}
%<3.2>
{\rm i}\hbar\frac{\partial\phi({\bm x},t)}{\partial
t}=\sqrt{m^2c^4-\hbar^2c^2\Delta}\,\phi({\bm x},t),
\end{equation}
so we apply the nonrelativistic formula for the Wigner function but we use 
the relativistic dynamics.  We point out that the similar form of the 
relativistic Wigner function has been already utilized in \cite{10} and \cite{11} for 
the Dirac spin one-half particle.  Evidently, we can also express the Wigner 
function in terms of the momentum representation
\begin{equation}
%<3.3>
W({\bm x},{\bm p},t)=\frac{1}{(2\pi)^3\hbar^6}\int d^3{\bm k}\tilde\phi^*({\bm p}-{\bm k}/2,t)
\tilde\phi({\bm p}+{\bm k}/2,t)e^{{\rm i}\frac{{\bm k}\mbox{\boldmath$\scriptstyle{\cdot}$}{\bm x}}{\hbar}},
\end{equation}
where $\tilde\phi({\bm p},t)$ fulfills (2.3).  It is also clear the Wigner 
function given by (3.1) or equivalently (3.3) can be immediately 
generalized to the case of of a particle in a potential field by 
postulating that evolution of states is described by (2.14) and (2.12), 
respectively.

By differentiating both sides of (3.3) with respect to time and using (2.3) as well as the 
identity
\begin{equation}
%<3.4>
\int d^3{\bm q}W({\bm q},{\bm p},t)e^{-{\rm i}\frac{{\bm q}
\mbox{\boldmath$\scriptstyle{\cdot}$}{\bm 
k}}{\hbar}}=\frac{1}{\hbar^3}\tilde\phi^*({\bm p}-{\bm 
k}/2,t)\tilde\phi({\bm p}+{\bm k}/2,t),
\end{equation}
we find
\begin{align}
%<3.5>
&\frac{\partial W({\bm x},{\bm p},t)}{\partial 
t}=-\frac{2c^2}{(2\pi\hbar)^3}\nonumber\\
&\times{\bm p}{\bm\cdot}{\bm\nabla}\int d^3{\bm k}d^3{\bm q}
\frac{W({\bm q},{\bm p},t)e^{{\rm i}\frac{{\bm k}
\mbox{\boldmath$\scriptstyle{\cdot}$}({\bm x}-{\bm q})}{\hbar}}}
{\sqrt{({\bm p}-{\bm k}/2)^2c^2+m^2c^4}+\sqrt{({\bm p}+{\bm k}/2)^2c^2+m^2c^4}}.
\end{align}
It thus appears that in opposition to (2.9) the evolution of the Wigner 
function (3.1) in the case of a free particle is nonlocal.  One finds easily that 
in the limit $c\to\infty$ (3.5) reduces to (2.8) that is we get the correct 
nonrelativistic evolution.  Finally, putting in (3.5) $m=0$ we obtain the 
massless limit such that
\begin{equation}
%<3.6>
\frac{\partial W({\bm x},{\bm p},t)}{\partial t}=-\frac{2c}{(2\pi\hbar)^3}
{\bm p}{\bm\cdot}{\bm\nabla}\int d^3{\bm k}d^3{\bm q}
\frac{W({\bm q},{\bm p},t)e^{{\rm i}\frac{{\bm k}
\mbox{\boldmath$\scriptstyle{\cdot}$}({\bm x}-{\bm q})}{\hbar}}}
{|{\bm p}-{\bm k}/2|+|{\bm p}+{\bm k}/2|}\qquad (m=0).
\end{equation}
where $|{\bm a}|$ designates the norm of the vector ${\bm a}$.  The 
formula (3.5) enables to easily demonstrate the correctness of the 
nonrelativistic limit, nevertheless we have the simpler relation 
describing the evolution of the Wigner function that can be derived with 
the use of the identity \cite{12}
\begin{equation}
%<3.7>
\int_{-\infty}^\infty\sqrt{x^2+a^2}\,e^{{\rm i}px}\,dx =
-\frac{2a}{|p|}K_1(a|p|)
\end{equation}
and the differentiation rule satisfied by the Bessel functions of the form
\begin{equation}
%<3.8>
K'_1(z)=\frac{1}{z}K_1(z)-K_2(z).
\end{equation}
Namely, we have
\begin{equation}
%<3.9>
\frac{\partial W({\bm x},{\bm p},t)}{\partial t}=\frac{2m^2c^3}{(2\pi\hbar)^2}
\int d^3{\bm q}\frac{K_2\left(\frac{2mc}{\hbar}|{\bm x}-{\bm q}|\right)}{|{\bm x}-{\bm q}|^2}
\sin\frac{2{\bm p}{\bm\cdot}({\bm x}-{\bm q})}{\hbar}W({\bm q},{\bm p},t).
\end{equation}
We remark that (3.9) has the form analogous to the Salpeter equation for a 
free particle written in the form of the integro-differential equation 
\cite{6}.  On taking the limit $m\to0$ of Eq.\ (3.9) and using the
asymptotic formula
\begin{equation}
%<3.10>
K_2(z) = \frac{2}{z^2},\qquad z\to 0,
\end{equation}
we arrive at the following massless limit of Eq.\ (3.9)
\begin{equation}
%<3.11>
\frac{\partial W({\bm x},{\bm p},t)}{\partial t}=\frac{c}{(2\pi)^2}
\int d^3{\bm q}\frac{\sin\frac{2{\bm p}{\bm\cdot}({\bm x}-{\bm q})}{\hbar}}{|{\bm x}-{\bm q}|^4}
W({\bm q},{\bm p},t).
\end{equation}
The one-dimensional version of the formula (3.9) corresponding to the case 
of a relativistic particle moving in a line is
\begin{equation}
%<3.12>
\frac{\partial W(x,p,t)}{\partial t}=\frac{2mc^2}{\pi\hbar}
\int_{-\infty}^{\infty} dq\frac{K_1\left(\frac{2mc}{\hbar}|x-q|\right)}{|x-q|}
\sin\frac{2p(x-q)}{\hbar}W(q,p,t).
\end{equation}
An immediate consequence of (3.12) and the asymptotic formula (2.55) is 
the following limit $m=0$ of (3.12)
\begin{equation}
%<3.13>
\frac{\partial W(x,p,t)}{\partial t}=\frac{c}{\pi}
\int_{-\infty}^{\infty} dq\frac{\sin\frac{2p(x-q)}{\hbar}}{(x-q)^2}W(q,p,t)\qquad (m=0).
\end{equation}

In order to illustrate the approach introduced in this section based on 
the definition of the Wigner function (3.1) and compare it with the 
Zavialov-Malokostov formalism discussed in the previous section we now 
consider a free particle on a line and the states (2.28) and (2.54) 
referring to the massless and massive particle, respectively.  The 
relativistic Wigner function for the motion in a line takes the form
\begin{equation}
%<3.14>
W(x,p,t)=\frac{1}{2\pi\hbar^2}\int_{-\infty}^{\infty}dk\,\tilde\phi^*(p-k/2,t)
\tilde\phi(p+k/2,t)e^{\frac{{\rm i}kx}{\hbar}}, 
\end{equation}
where $\tilde\phi(p,t)$ satisfies (2.21) and (2.26) for the massive and 
massless case, respectively.  Consider first the massless particle and the 
state (2.28).  On inserting (2.28) into (3.14) with $\hbar=1$ and making 
use of the identity \cite{8}
\begin{equation}
%<3.15>
\int_0^\infty 
e^{-px}\cos(qx+\lambda)dx=\frac{1}{p^2+q^2}(p\cos\lambda-q\sin\lambda),\qquad 
p>0,
\end{equation}
we arrive at the following formula for the Wigner function
\begin{align}
%<3.16>
W(x,p,t)=&\frac{a}{\pi}e^{-2a|p|}\left\{|p|\frac{\sin2(pt-x|p|)}{pt-x|p|}\right.\nonumber\\
&{}+\left.\frac{a}{a^2+x^2}\left[\cos2(pt-x|p|)+\frac{x}{a}\cos2(pt-x|p|)\right]\right\}.
\end{align}
The time evolution of the Wigner function (3.16) is demonstrated in Fig.\ 
4 and Fig.\ 5.  The characteristic feature of the Wigner function is the existence of 
the stable global maximum centered around $x=0$, $p=0$, and appearance of 
local extrema whose number increases as the time develops.
\begin{figure*}
%<figure 4>
\centering
\begin{tabular}{c@{}c}
\includegraphics[width =.5\textwidth]{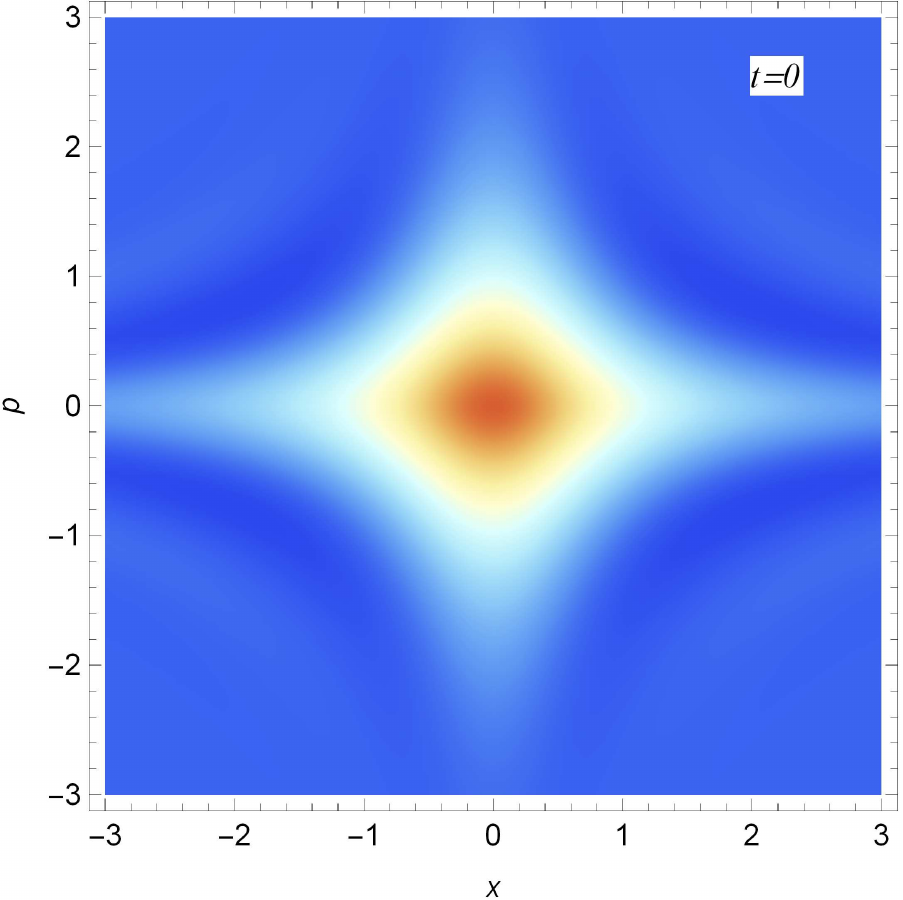}&
\includegraphics[width =.5\textwidth]{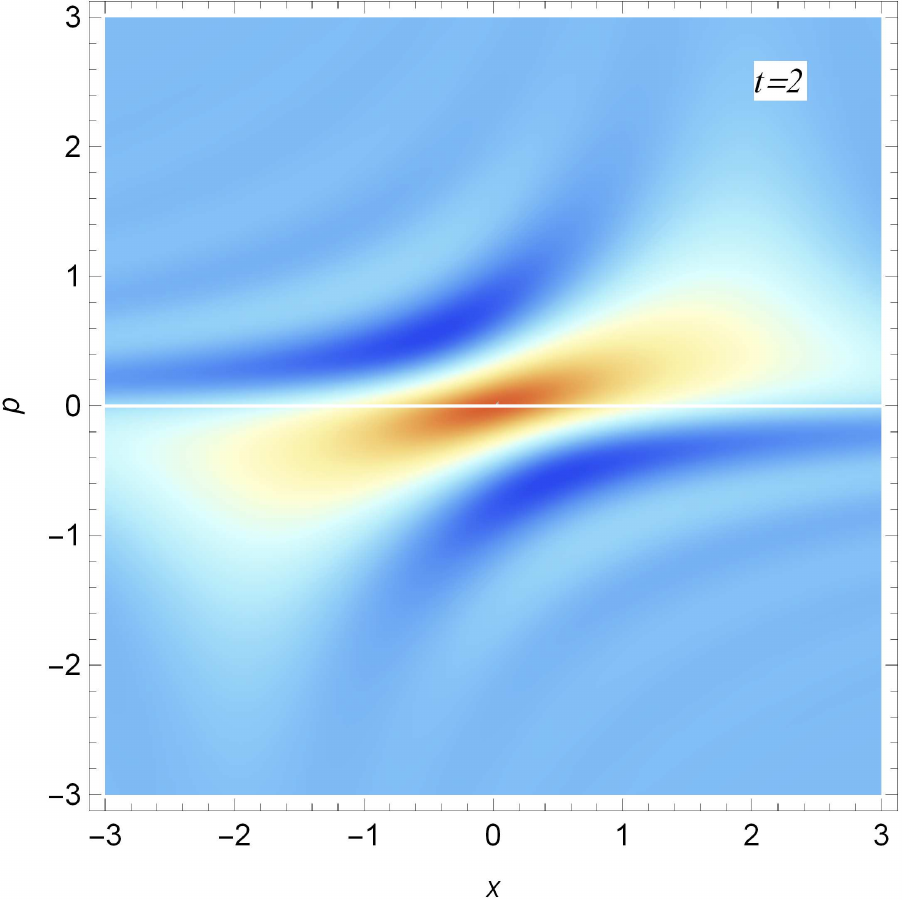}\\
\includegraphics[width =.5\textwidth]{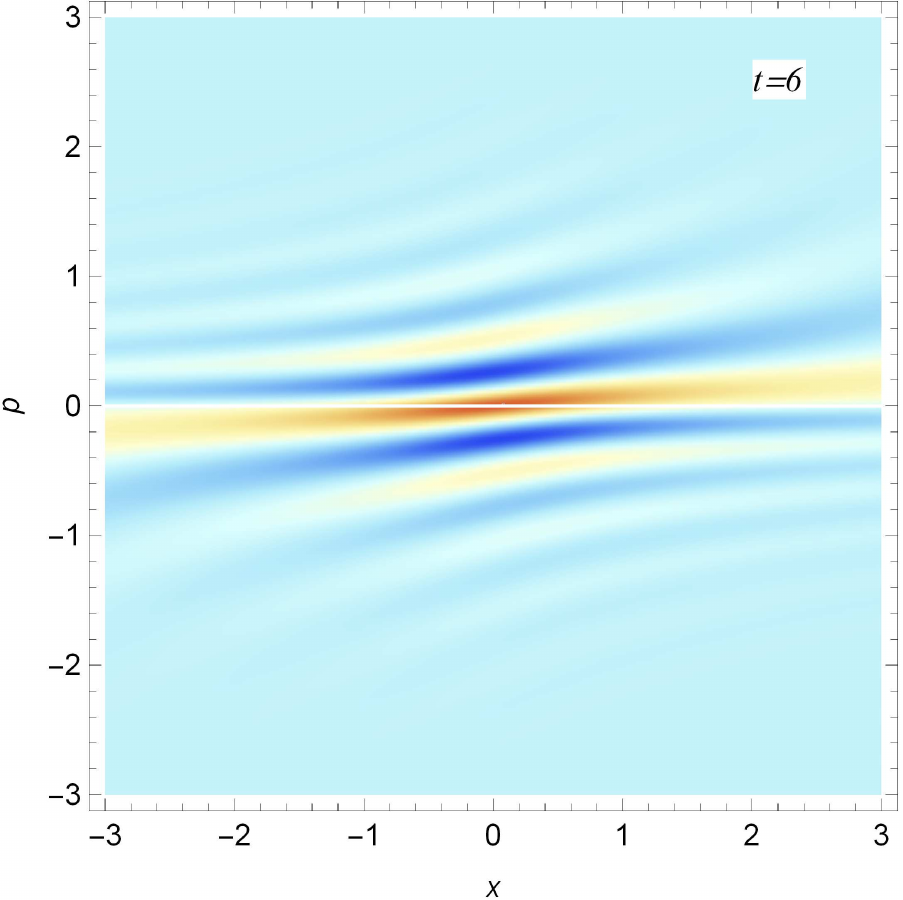}&
\includegraphics[width =.5\textwidth]{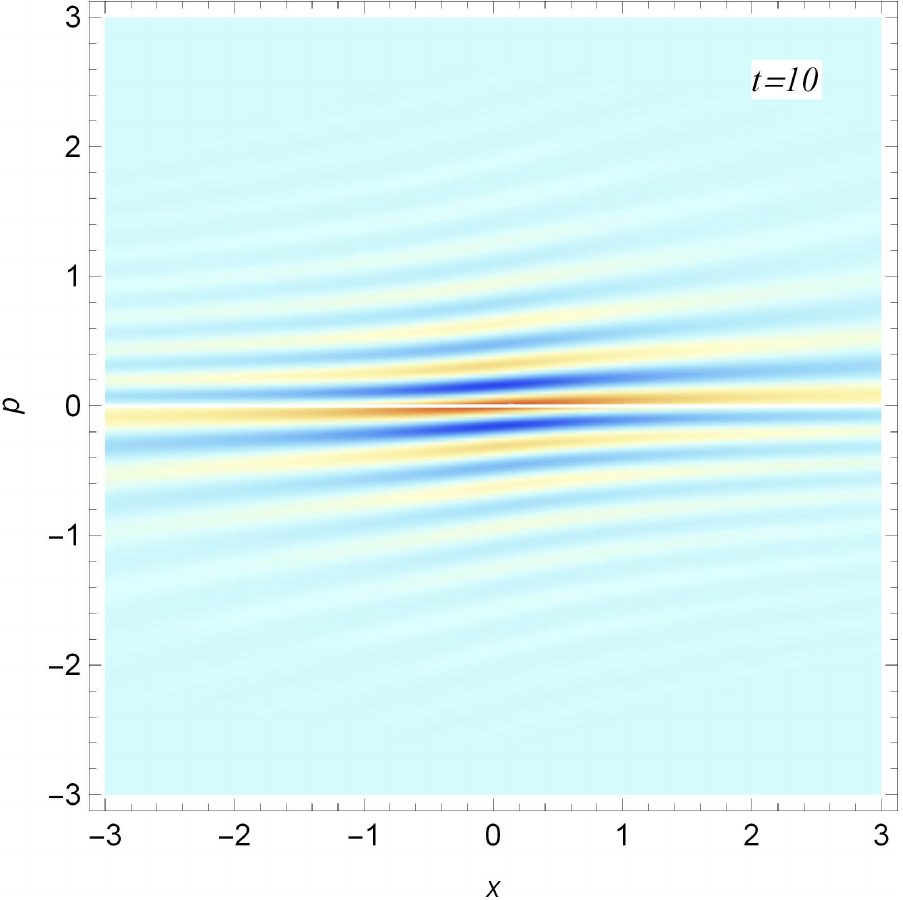}
\end{tabular}
\caption{The behavior of the Wigner function (3.16) corresponding to the case
of the free massless particle described by the solution (2.31). The parameters 
have the same values as in Fig.\ 1.  For the details of the figure bottom right 
see Fig.\ 5.}
\end{figure*}
\begin{figure*}
%<figure 5>
\centering
\includegraphics[scale=1.5]{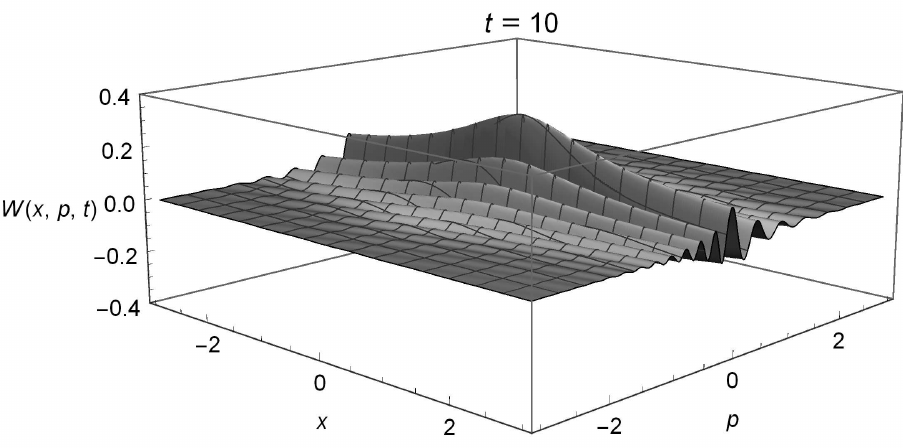}
\caption{Figure 4 bottom right shown in 3D presentation.  The multiple local 
extrema arising as time develops and the non-positivity of the Wigner function are 
easily seen.  The central global maximum is stable.}
\end{figure*}

We finally discuss the case of a free massive particle on a line and the 
state (2.54).  Substituting (2.54) into (3.14) and setting $\hbar=1$ we 
get
\begin{align}
%<3.17>
W(x,p,t)=&\frac{1}{2m\pi K_1(2ma)}\nonumber\\
&\times\int_0^\infty dk\exp\{-a[\sqrt{(p-k/2)^2+m^2}
+\sqrt{(p+k/2)^2+m^2}]\}\nonumber\\
&\times\cos\{t[\sqrt{(p-k/2)^2+m^2}-\sqrt{(p+k/2)^2+m^2}]+kx\}.
\end{align}
The authors do not know any analytic expression for the integral from 
(3.17).  The only exception besides $m=0$ is the case $p=0$ when it 
reduces to the integral of the form \cite{8} (see also (2.49))
\begin{align}
%<3.18>
\int_0^\infty dx\,\exp(-\alpha\sqrt{x^2+\beta^2})\cos\gamma x =&
\frac{\alpha\beta}{\sqrt{\alpha^2+\gamma^2}}K_1(\beta\sqrt{\alpha^2+\gamma^2}),\nonumber\\
&\qquad{\rm Re}\alpha>0,\, {\rm Re}\beta >0.
\end{align}
The Wigner function is then given by
\begin{equation}
%<3.19>
W(x,0,t)=\frac{1}{\pi K_1(2ma)}\frac{a}{\sqrt{a^2+x^2}}K_1(2m\sqrt{a^2+x^2}).
\end{equation} 
The time development of the Wigner function (3.17) obtained by numerical 
calculation of the integral is shown in Fig.\ 6 and Fig.\ 7.  In opposition to the 
massless case the global maximum is not stable in the limit of large $t$ and 
behaves similarly as in the case of (2.59) that is preserves maximal value 
but flatten.  Analogously as in the massless case we have also multiple 
local extrema whose number increases with time.  As with (2.59) the Wigner function 
can take the negative values as well (see Fig.\ 7).
\begin{figure*}
%<figure 6>
\centering
\begin{tabular}{c@{}c}
\includegraphics[width =.5\textwidth]{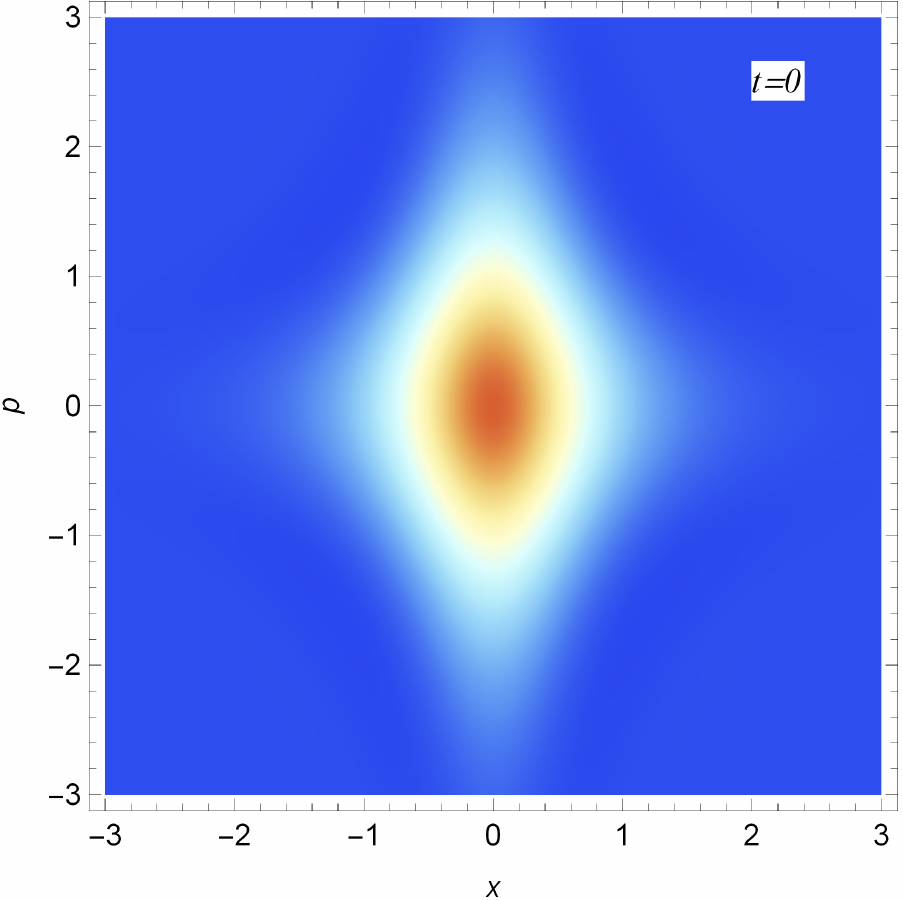}&
\includegraphics[width =.5\textwidth]{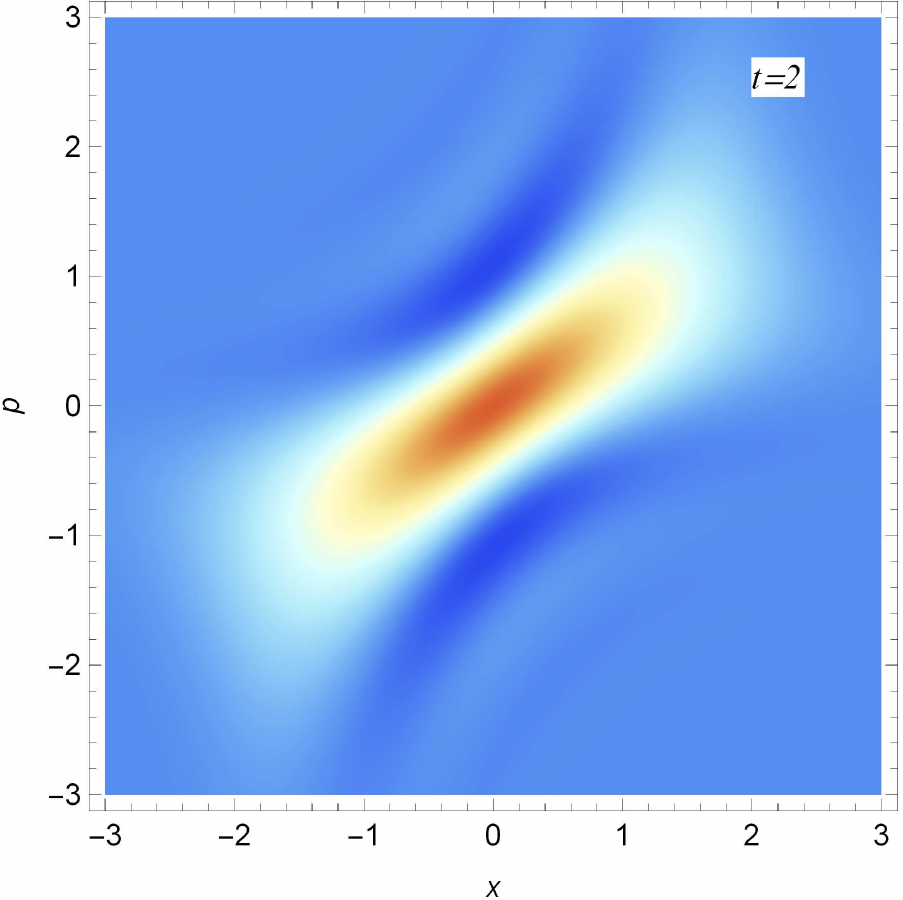}\\
\includegraphics[width =.5\textwidth]{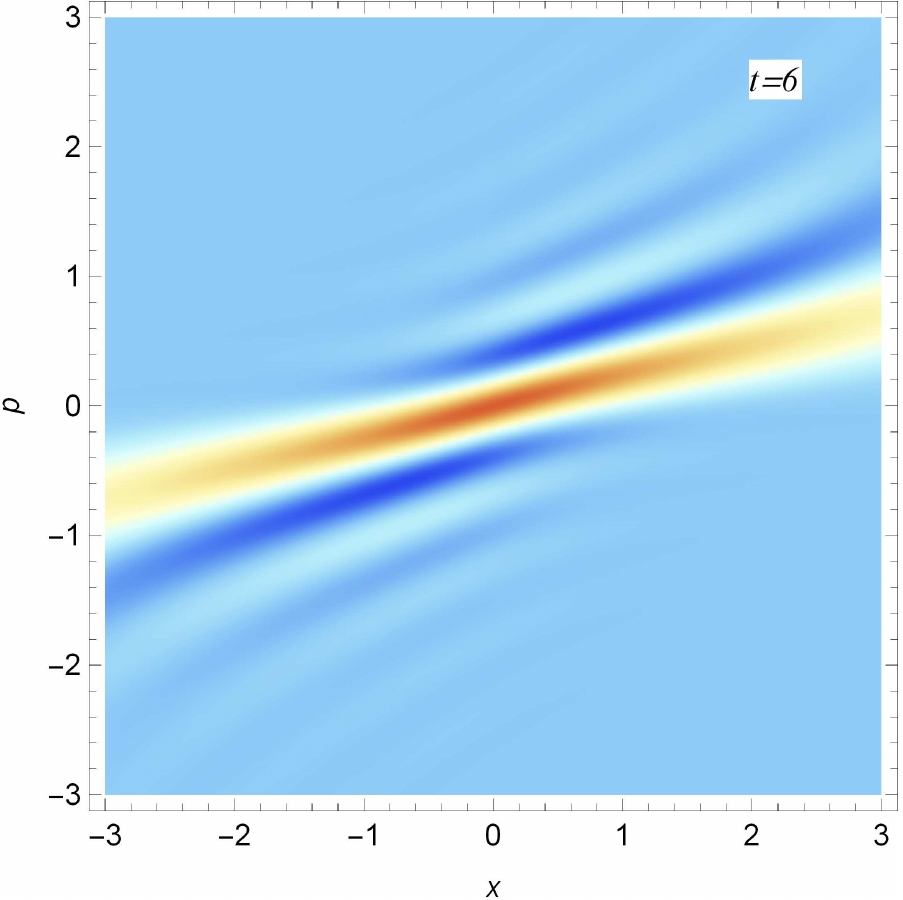}&
\includegraphics[width =.5\textwidth]{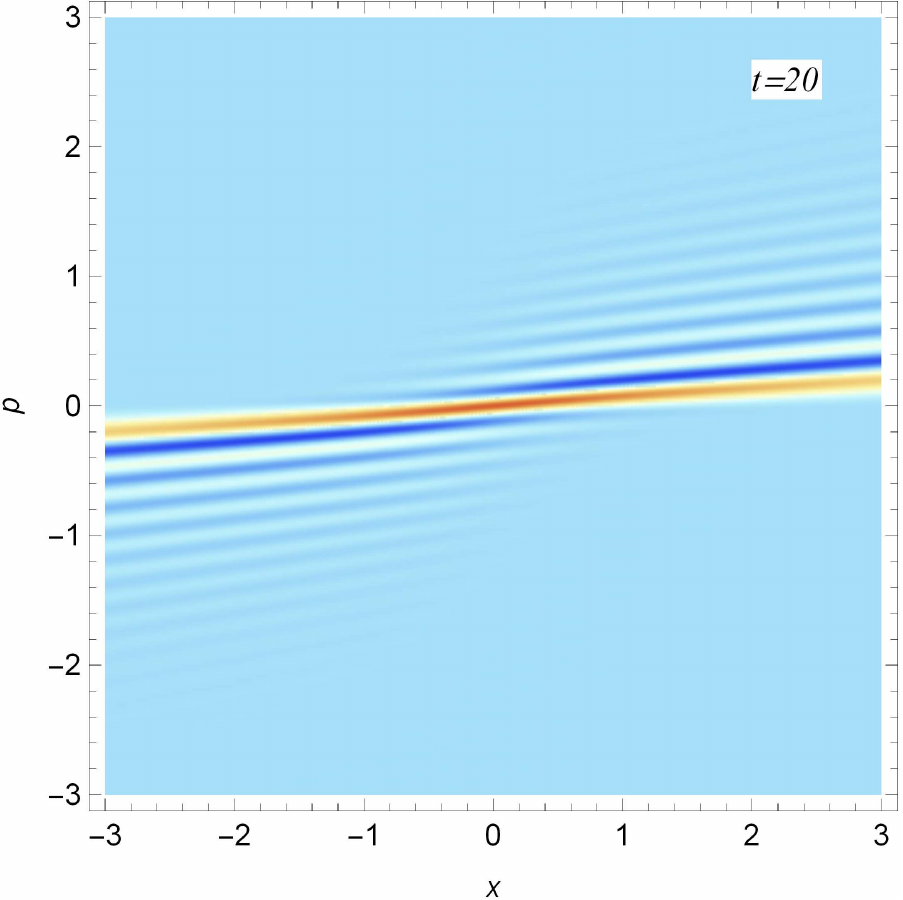}
\end{tabular}
\caption{The time evolution of the Wigner function (3.17) referring to free 
massive particle in the state (2.54) the same as in Fig.\ 2. The values of the 
parameters $m=1$ and $a=1$ are the same as well.  The details of the figure 
bottom right are shown in Fig.\ 7.}
\end{figure*}
\begin{figure*}
%<figure 7>
\centering
\includegraphics[scale=1.3]{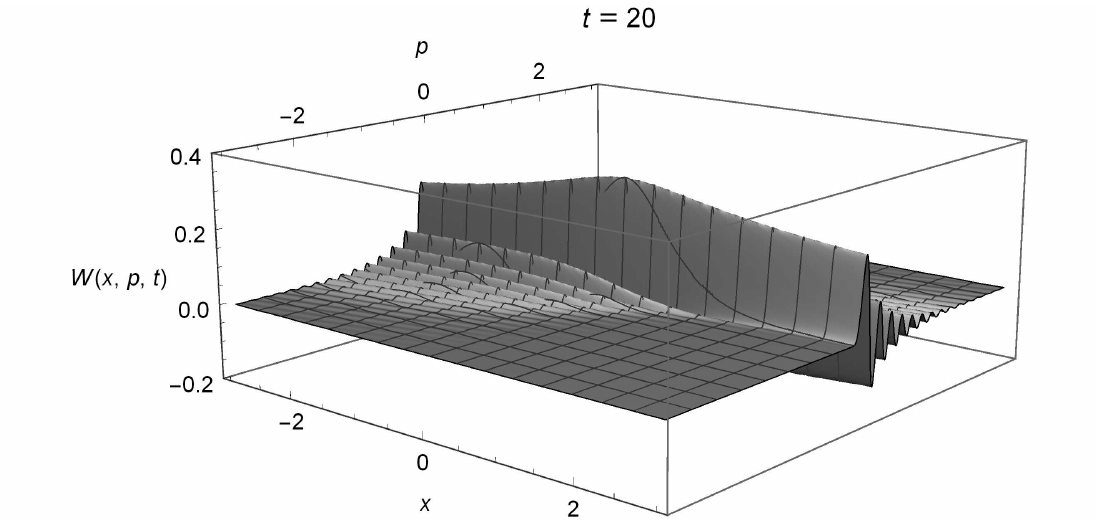}
\caption{The 3D version of Fig.\ 6 bottom right.  Analogously as for $m=0$ 
(see Fig.\ 3) the Wigner function has multiple local extrema. The regions where
the Wigner function takes the negative values are also easily observed.}
\end{figure*}
\section{Summary}
In this work we discuss two approaches to relativistic generalization of 
the Wigner function for a free spinless particle based on the Salpeter 
equation.  The first one introduced by Zavialov and Malokostov postulates 
the local law of evolution for the Wigner function in the case of a free 
particle. Its flaw is the problematic behavior of the Wigner function in 
the limit $m=0$.  The problems with the limit $m=0$ are related to the 
definition of the Wigner function (2.9) utilizing without clear physical 
motivation, the ``sum'' of two fourvectors on the mass hyperboloid that 
remains on the hyperboloid.  Such ``sum'' cannot be in general defined on 
a cone in the case of massless particles.  On the other hand, we have 
demonstrated that the difficulties with the discussed formalism are 
connected with the simultaneous validity of (2.7) and (2.13).  Therefore, 
the question naturally arises as to whether the Wigner function satisfying 
assumptions (2.4), (2.5) and (2.7) is unique.  The second approach is based 
on the standard definition applied in nonrelativistic quantum mechanics 
but with states evolving according to the relativistic Salpeter equation.
In opposition to (2.7) assumed by Zavialov and Malokostov, the dynamics of 
such Wigner function is nonlocal in the case of a free particle.  On the 
one hand this is quite plausible in view of the nonlocality of the Salpeter 
equation related only to the kinetic energy term \cite{6}.  On the other 
hand, we know for example that in spite of the nonlocality of the current
(2.16), it satisfies the local continuity equation (2.15).  Bearing in 
mind all pros and cons we find that there is no conclusive evidence which
candidate for the relativistic Wigner function is better.   Nonetheless,
it seems to us that the observations obtained herein, especially the analytic 
expressions for the Wigner functions, would be of importance for the further 
studies of the subject.  Finally, we point out possible application of the method 
for construction of the Wigner functions for general Lie groups developed in \cite{13,14,15}.  
Indeed, the role of addition of vectors on the mass hyperboloid in the 
Zavialov-Malokostov approach to the relativistic Wigner function and results 
obtained with the help of the method in the nonrelativistic case suggest some 
group-theoretic context of the problem.  Nevertheless, in spite of the fact that 
Wigner functions obtained by means of the method have many desirable properties 
such as for example the covariance, their physical interpretation can be 
unclear (see for instance \cite{16}).  For this reason, we defer this issue to 
our future work.
\section*{Acknowledgements}
This work has been supported by the Polish National Science Centre under Contract
2014/15/B/ST2/00117 and by the University of Lodz.

\end{document}